\documentclass[prl, twocolumn, showpacs, nofootinbib, amsmath,amssymb, floatfix, eqsecnum]{revtex4-1}
\pdfoutput=1
\usepackage{amsmath}
\usepackage{amssymb}
\usepackage{amsthm}
\usepackage{amsfonts}
\usepackage{comment}%https://www.overleaf.com/project/603d67a5abedda342ae4c069
\usepackage[normalem]{ulem}
\usepackage{graphicx}
\usepackage{color,framed}
\usepackage{hyperref}
\usepackage{times}
\usepackage{enumerate}
\usepackage{lipsum}
\usepackage{slashed}
\usepackage{url}
\usepackage{bbm}
\usepackage{chngcntr}
\usepackage{tikz}
\usetikzlibrary{spy}
\usetikzlibrary{patterns}
\usetikzlibrary{calc,arrows,positioning}
\usetikzlibrary{arrows,shadows}
\usetikzlibrary{decorations.pathmorphing}	% For Feynman 
\usetikzlibrary{decorations.markings}
\usepackage{pgfplots}
\usepackage{pgfplotstable}
\counterwithout{equation}{section}

\hypersetup{
    colorlinks=true, %set true if you want colored links
    linktoc=all,     %set to all if you want both sections and subsections linked
    linkcolor=blue,  %choose some color if you want links to stand out
}

\def \beq {\begin{equation}}
\def \eeq {\end{equation}}
\def \beqa {\begin{eqnarray}}
\def \eeqa {\end{eqnarray}}
\def \bseq {\begin{subequations}}
\def \eseq {\end{subequations}}

\begin{document}

\title{Antiunitary symmetry breaking and a hierarchy of purification transitions in Floquet non-unitary circuits}

\author{Carolyn Zhang}
\affiliation{Department of Physics, Kadanoff Center for Theoretical Physics, University of Chicago, Chicago, Illinois 60637,  USA}
\affiliation{Department of Physics, Harvard University, Cambridge, MA02138, USA}
\author{Etienne Granet}
\affiliation{Quantinuum, Leopoldstrasse 180, 80804 Munich, Germany}
\date{\today}

\begin{abstract}
We consider how a maximally mixed state evolves under $(1+1)D$ Floquet non-unitary circuits with an antiunitary symmetry that squares to identity, that serves as a generalized $\mathcal{PT}$ symmetry. %It is expected that in generic interacting quantum systems, the symmetry breaking point of this antiunitary symmetry corresponds to a purification transition for mixed states. Specifically, the symmetric phase is identified with a phase that does not purify (mixed phase) and the symmetry breaking phase is identified with a phase that purifies in a time that is independent of the system size (strongly purifying phase). 
Upon tuning a parameter, the effective Hamiltonian of the Floquet operator demonstrates a symmetry breaking transition. We show that this symmetry breaking transition coincides with different kinds of purification transitions. Gaussian non-unitary circuits are mixed (not purifying) on both sides of the symmetry breaking transition, while interacting but integrable non-unitary circuits are mixed on the symmetric side and ``weakly purifying" on the symmetry breaking side. In the weakly purifying phase, the initial mixed state purifies on a time scale proportional to the system size. We obtain numerically the critical exponents associated with the divergence of the purification time at the purification transition, which depend continuously on the parameters of the model. Upon adding a symmetric perturbation that breaks integrability, the weakly purifying phase becomes strongly purifying, purifying in a time independent of the system size, for sufficiently large system size. Our models have an extra $U(1)$ symmetry that divides the Hilbert space into different magnetization sectors, some of which demonstrate logarithmic scaling of entanglement in the weakly purifying phase.%We also obtain various entanglement transitions of pure states, that depend on the purification transition and the magnetization sector. 
\end{abstract}

\maketitle
\textbf{\emph{Introduction}}---While unitary evolution preserves the purity of a mixed state, non-unitary processes such as measurement and dissipation can steer a mixed state toward a pure state. In hybrid circuits built out of both unitary gates and projective measurement, it was shown that the volume-law to area-law entanglement transition for pure initial states\cite{skinner2019,li2018,li2019,chan2019} becomes a purification transition when the initial state is instead maximally mixed\cite{gullans2020,li2021error,ippoliti2021,biella2021,lunt2021,ippoliti2021measurement,altland2022,agrawal2022,block2022,potter2022,loio2023}. In particular, the volume-law phase corresponds to the ``mixed" phase, where the initial mixed state remains mixed up to time scales exponential in the system size, while the area-law phase corresponds to a strongly purifying phase, where the initial mixed state becomes pure in a time scale independent of the system size. Purification transitions also occur in more general non-unitary systems\cite{ashida2017,gopalakrishnan2021entanglement,bentsen2021,jian2021,turkeshi2022,kawabata2023}, with weak/forced measurements enabled by postselection.
% \CZ{i reworded this and removed markovian bath bc markovian bath is more general Lindbladian evolution}
 The volume-law phase in these systems correspond to a stronger mixed phase, where the system never purifies\cite{gopalakrishnan2021entanglement,fidkowski2021}.

\begin{figure}[tb]
   \centering
\includegraphics[width=.84\columnwidth]{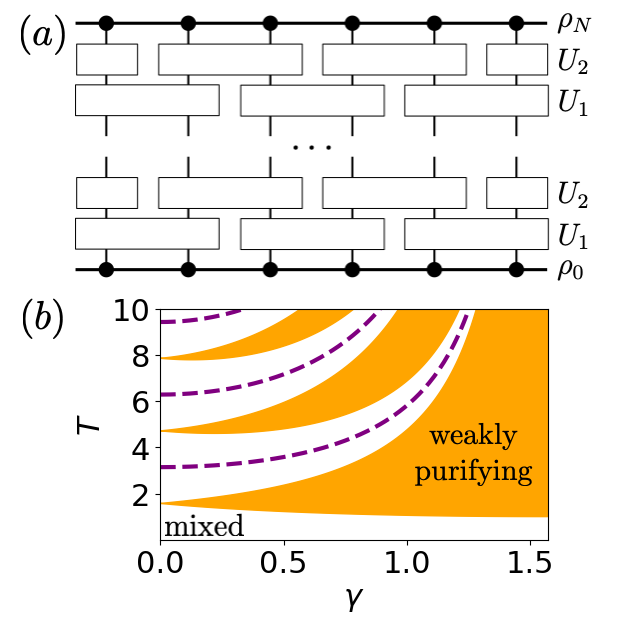} 
   \caption{$(a)$ The Floquet operator we consider consists of two layers of disjoint two-site non-unitary gates. We study the long-time density matrix $\rho_N$ after $N$ applications of this circuit. $(b)$ The phase diagram consists of mixed (white) and weakly purifying (orange) phases. The circuit is periodic in the parameter $T$, with periods separated by the dashed purple lines.}
   \label{phasediag}
\end{figure}
In this work, we study purification transitions in a family of integrable, Floquet non-unitary circuits in one dimension, that was first described in Ref.~\onlinecite{miao2022}. The circuits consist of repeated application of a Floquet operator $U_F$ built out of non-unitary gates. Importantly, $U_F$ respects an antiunitary symmetry $\mathcal{A}$ that squares to identity. This symmetry ensures that the purification transitions occurs at a nonvanishing value of a parameter $T$ that tunes the non-unitarity\cite{bender1998}. %Second, it has a $U(1)$ symmetry that separates the Hilbert space into magnetization sectors. This $U(1)$ symmetry does not affect the purification transitions, but does make the steady state entanglement depend on 

We find a hierarchy of different kinds of purification transitions in this family of circuits. In a certain parameter regime, the circuit becomes Gaussian, and demonstrates a mixed phase on both sides of the symmetry breaking transition. This result is in agreement with Ref.~\onlinecite{legal2022}. The interacting but integrable circuits demonstrate mixed to weakly purifying transitions, where in the latter the initial maximally mixed state purifies on a time scale linear in the system size. Even for finite system size, the purification time diverges as $T$ approaches the transition to the mixed phase. The exponents governing this divergence, in the thermodynamic limit, are numerically observed to be close to the critical exponents of an equilibrium integrable system related to our circuit \cite{luther1975,lukyanov1999correlation}.
% closely resemble certain exponents of equilibrium Hermitian systems\CZ{add ref}. 
Finally, upon adding symmetric perturbations that break integrability, the circuit demonstrates a mixed to strongly purifying transition, in agreement with Ref.~\onlinecite{gopalakrishnan2021entanglement}. These purification transitions also lead to entanglement transitions of individual quantum trajectories. For example, the mixed to weakly purifying transition gives a numerically observed volume-law to log-law entanglement transition in interacting but integrable systems, in certain total magnetization sectors.

We study these purification phases and transitions using analytic and numerical results on the spectrum of $U_F$. In particular, the purification time scale $\tau$ is related to the spectrum of $U_F$ by $\tau\sim\frac{1}{\Delta}$, where $\Delta=\log\left|\frac{\lambda_1}{\lambda_2}\right|$. Here, $\lambda_1$ and $\lambda_2$ are the eigenvalues of $U_F$ with the largest and second largest modulus respectively. Note that if $U_F$ is unitary, all of its eigenvalues have modulus 1, so $\Delta=0$ and $\tau=\infty$, meaning it must realize a mixed phase. Because the family of non-unitary circuits we study is integrable, we can obtain results on the spectrum of $U_F$ using Bethe ansatz techniques. 

\textbf{\emph{Purification transitions and antiunitary symmetries}}---
We begin with a review of some definitions and concepts relevant to our results. We define the purity of a density matrix $\rho$ as
\begin{equation}\label{purity}
\Pi(\rho)=\frac{\mathrm{Tr}(\rho^2)}{\mathrm{Tr}(\rho)^2}.
\end{equation}

In the following, we will begin with the maximally mixed state $\rho_0=\frac{1}{2^L}\mathbbm{1}$ and study the purity of $\rho_N$, where
\begin{equation}
\rho_N=\left(U_F\right)^N\rho_0\left(U_F^\dagger\right)^N=\frac{1}{2^L}\left(U_F\right)^N\left(U_F^\dagger\right)^N,
\end{equation}
as $N\to\infty$. Note that $\mathrm{Tr}(\rho_N)\neq 1$ in general, because $U_F$ is non-unitary. An important property of $U_F$ is that it respects an antiunitary symmetry $\mathcal{A}$: 
\begin{equation}\label{antiunitary}
\mathcal{A}U_F\mathcal{A}^{-1}=U_F^{-1}\qquad \mathcal{A}^2=\mathbbm{1}.
\end{equation}

$\mathcal{A}$ serves as a generalized $\mathcal{P}\mathcal{T}$ (parity-time reversal) symmetry. Specifically, it follows from (\ref{antiunitary}) that the effective non-Hermitian Hamiltonian $H_{\mathrm{eff}}$, given by $U=e^{-iH_{\mathrm{eff}}}$, commutes with $\mathcal{A}$. As a result, the eigenvalues of $H_{\mathrm{eff}}$ come in complex conjugate pairs. It is well-known that this condition implies that $H_{\mathrm{eff}}$ can have eigenvalues that are all real, even if it is not Hermitian\footnote{Note that for this argument to work, we must assume that $H_{\mathrm{eff}}$ does not have any degenerate eigenvalues. This is compatible with the fact that $\mathcal{A}^2=1$; if $\mathcal{A}^2=-1$, then $H_{\mathrm{eff}}$ would have Kramers degeneracies.}. Let us consider the case where  $H_{\mathrm{eff}}$ can be written as
\begin{equation}
H_{\mathrm{eff}}=H_0+\lambda H_1\,,
\end{equation}
where $H_0$ is Hermitian and $H_1$ is non-Hermitian, and with $\lambda$ a parameter tuning the non-hermiticity of the model. $H_0$ has real and generically non-degenerate eigenvalues, and for small $\lambda$, the eigenvalues of $H_{\mathrm{eff}}$ are perturbatively close to those of $H_0$. The constraint that the eigenvalues of $H_{\mathrm{eff}}$ must come in complex conjugate pairs then implies that, unless the eigenvalues of $H_0$ are degenerate, the spectrum of $H_{\mathrm{eff}}$ must remain real. Therefore, for small $\lambda$, the eigenvalues of $U_F=e^{-iH_{\mathrm{eff}}}$ still all have unit modulus, so $\Delta=0$. 

The spectrum of $U_F$ changes suddenly at a critical value of $\lambda$ where the system ``spontaneously breaks" $\mathcal{A}$. At this point, at least two eigenvalues of $H_{\mathrm{eff}}$ meet on the real axis and split into the complex plane. In the symmetry breaking phase, the eigenvalues of $H_{\mathrm{eff}}$ are no longer all real; some are complex and come in complex conjugate pairs. If there is a unique eigenvalue of $H_{\mathrm{eff}}$ with largest positive imaginary part, then $U_F$ has a unique eigenvalue with largest modulus, leading to a nonzero $\Delta$. $\left(U_F\right)^N$ as $N\to\infty$ projects onto the eigenstate corresponding to this eigenvalue, so the Floquet system takes an initial mixed state to a pure state. 

Note that the antiunitary symmetry $\mathcal{A}$ ensures that there is an entire mixed \emph{phase}. Without it, a generic perturbation $\lambda H_1$ would immediately make the eigenvalues of $H_{\mathrm{eff}}$ complex, resulting in a purifying phase. We will show that indeed our circuit has such a symmetry, and that breaking the symmetry explicitly removes the mixed phase.

\textbf{\emph{Integrable circuit}}---
We will study the integrable circuit presented in Ref.~\onlinecite{miao2022}. It consists of two layers of disjoint two-site gates:
\begin{equation}\label{model1}
U_F=U_2U_1\,,
\end{equation}
where 
\begin{align}
\begin{split}\label{udef}
U_1&=\prod_{m=1}^{L/2}\mathrm{exp}(-ih_{2m,2m+1}T)\\
U_2&=\prod_{m=1}^{L/2}\mathrm{exp}(-ih_{2m-1,2m}T)\,,
\end{split}
\end{align}
and
\begin{align}
\begin{split}\label{hdef}
h_{m,n}&=-\frac{1}{2}\left(\sigma^x_m\sigma^x_n+\sigma^y_m\sigma^y_n+\cos\gamma \sigma^z_m\sigma^z_n\right)\\
&+\cos\gamma-\frac{i}{2}\left(\sigma^z_m-\sigma^z_n\right)\sin\gamma\,.
\end{split}
\end{align}

$T$ and $\gamma$ are real parameters that tune through different purification phases; the circuit remains integrable for all values of $T$ and $\gamma$. The lines $\gamma=0$ and $\gamma=\frac{\pi}{2}$ are particularly simple to analyze. At the former, $U_F$ describes unitary time evolution, so the circuit gives a mixed phase for all values of $T$. At the latter, the circuit is Gaussian and non-unitary. We will show that along this line, the circuit is also mixed for all values of $T$. Between these two limits, the circuit is interacting and non-unitary, and demonstrates mixed to weakly purifying transitions. 

To simplify the analysis, note that the circuit is periodic in $T$:
\begin{align}
\begin{split}
\mathrm{exp}&(-ih_{m,m+1}T)\\
&=\mathbbm{1}+\frac{1}{2\cos\gamma}\left(\mathrm{exp}(2iT\cos\gamma)-1\right)h_{m,m+1}\,.
\end{split}
\end{align}
Therefore, we only need to study values of $T$ between $0$ and $\frac{\pi}{\cos\gamma}$. In the following, we will restrict to $T$ in this range, with the understanding that the phase structure we obtain is periodic in $T$ with period $\frac{\pi}{\cos\gamma}$.

$U_F$ has an antiunitary symmetry $\mathcal{A}$ and a $U(1)$ symmetry $U_{\theta}$. These symmetries are given explicitly by
\begin{align}
\begin{split}
  \mathcal{A}&=S\mathcal{P}\mathcal{T}\\
          U_{\theta}&=\mathrm{exp}(i\theta\sum_m\sigma^z_m).
    \end{split}
    \end{align}
    
Here, $S$ is a unit shift (translation) operator, $\mathcal{P}$ is a spatial parity operator, and $\mathcal{T}$ performs complex conjugation. Specifically,
\begin{align}
\begin{split}
\mathcal{P}h_{m,m+1}\mathcal{P}^{-1}&=h_{L-m,L-m+1}\\
\mathcal{T} h_{m,m+1}\mathcal{T}^{-1}&=h_{m+1,m}\,.
\end{split}
\end{align}

One can check that $\mathcal{A}$ satisfies (\ref{antiunitary}) and $U_{\theta}$ commutes with $U_F$. The $U(1)$ symmetry means that if we start in (possibly mixed) state in a certain magnetization sector, then $U_F$ preserves the magnetization sector. We will only study the purification rate of a maximally mixed state, which includes all magnetization sectors. %However, it does affect the entanglement transitions of pure states.

It will be convenient to introduce a parameter $\alpha$ defined as
\begin{equation}
    \alpha=-\frac{1}{2}\log \frac{\cos(\gamma-T\cos\gamma)}{\cos(\gamma+T\cos\gamma)}\,.
\end{equation}

$\alpha$ detects whether or not $\mathcal{A}$ is broken. Specifically, $\alpha$ is real for $0<T<T^-_c$ and $T>T_c^+$ in the range $T\in\left[0,\frac{\pi}{\cos\gamma}\right)$, with
\begin{equation}
    T^\pm_c=\frac{\pi\pm 2\gamma}{2\cos\gamma}\,.
\end{equation}

It was shown in Ref.~\onlinecite{miao2022} that the eigenvalues of $U_F$ all have modulus 1 when $\alpha$ is real, so $\mathcal{A}$ is not broken in this regime. On the other hand, for $T_c^-<T<T_c^+$, $\alpha$ has imaginary part $\frac{\pi}{2}$. In this interval, the eigenvalues of $U_F$ no longer all have modulus 1, and $\mathcal{A}$ is broken.

As shown in Ref.~\onlinecite{miao2022}, this circuit is solvable with the Bethe ansatz. Eigenvalues $\Lambda$ of $U_F$ are given by
\begin{equation}
    \Lambda=\prod_{m=1}^M \frac{\sinh(\lambda_m+\frac{\alpha}{2}+\frac{i\gamma}{2})\sinh(\lambda_m-\frac{\alpha}{2}-\frac{i\gamma}{2})}{\sinh(\lambda_m+\frac{\alpha}{2}-\frac{i\gamma}{2})\sinh(\lambda_m-\frac{\alpha}{2}+\frac{i\gamma}{2})}\,.
\end{equation}
with the Bethe roots $\{\lambda_m\}$ satisfying the Bethe equations
\begin{equation}\label{be}
\begin{aligned}
    &\left( \frac{\sinh(\lambda_m+\frac{\alpha}{2}+\frac{i\gamma}{2})\sinh(\lambda_m-\frac{\alpha}{2}+\frac{i\gamma}{2})}{\sinh(\lambda_m+\frac{\alpha}{2}-\frac{i\gamma}{2})\sinh(\lambda_m-\frac{\alpha}{2}-\frac{i\gamma}{2})} \right)^{L/2}\\
    &\qquad\qquad=\prod_{n\neq m}^M \frac{\sinh(\lambda_m-\lambda_n+i\gamma)}{\sinh(\lambda_m-\lambda_n-i\gamma)}\,,
\end{aligned}
\end{equation}
where $M\in\left[1,L\right]$
% \CZ{changed from $[0,L-1]$ to $[1,L]$}
is an integer.

\textbf{\emph{Gaussian limit: volume-law phase}}---
We first consider the $\mathcal{A}$ symmetry broken phase given by $T\in(T_c^-,T_c^+)=(1,\infty)$ along the free fermion line at $\gamma=\frac{\pi}{2}$. We will show that the number of eigenvalues of $U_F$ with the same largest modulus for all such values of $T$ is exponentially large in $L$. This means that $\Delta=0$, so the circuit realizes a mixed phase on both sides of the $\mathcal{A}$ symmetry breaking transition. For an initial state that is pure, this implies that the long-time steady state is volume-law entangled for all $T$. This agrees with Ref.~\onlinecite{legal2022}, which studied a Gaussian circuit with $\mathcal{P}\mathcal{T}$ symmetry and observed a volume-law phase on both sides of the symmetry breaking transition.

For $\gamma=\frac{\pi}{2}$, the right hand side of \eqref{be} is $(-1)^{M-1}$, and its solutions are given by\cite{supp}
\begin{equation}
e^{2\lambda_k^{\pm}}=-i\tan(\tfrac{2\pi k}{L})\sinh\beta\pm\sqrt{1-\tan^2(\tfrac{2\pi k}{L})\sinh^2\beta}\,,
\end{equation}
where $\beta=\alpha-\frac{i\pi}{2}$ encodes the $T$ dependence and $k\in\left(-\frac{L}{4},\frac{L}{4}\right)$ taking integer/half-integer values for $M$ odd/even. Along this line, there are two kinds of roots $\lambda_k^s$, where $s\in\{+,-\}$. For $k$ satisfying $\tan^2(\tfrac{2\pi k}{L})\sinh^2\beta<1$, $\lambda_k^s$ is purely imaginary with absolute value smaller (resp. larger) than $\frac{\pi}{4}$ for $s=+$ (resp. $s=-$). We call these roots ``type (i) roots". For $k$ satisfying $\tan^2(\tfrac{2\pi k}{L})\sinh^2\beta>1$, the $\lambda_k^s$ is complex with imaginary part $\pm \frac{\pi}{4}$. We denote these by ``type (ii) roots". Each eigenvalue $\Lambda$ of $U_F$ corresponds to a subset $\mathcal{M}=\{\lambda_k^s\}$ of the roots, and is given by
\begin{equation}
\Lambda=\prod_{\lambda_k^s\in\mathcal{M}}\frac{\cosh(2\lambda_k^s)+\cosh\beta}{\cosh(2\lambda_k^s)-\cosh\beta}\,.
\end{equation}

We see that the type (ii) roots do not contribute to $|\Lambda|$. On the other hand, type (i) roots increase (resp. decrease) $|\Lambda|$ when smaller (resp. larger) than $\frac{\pi}{4}$ in modulus. Therefore, the largest $|\Lambda|$s are obtained by taking all type (i) roots $\{\lambda_k^+\}$ with modulus smaller than $\frac{\pi}{4}$, and any subset of the type (ii) roots. It follows that for any $T\in(1,\infty)$, there is an exponentially large number of eigenvalues of $U_F$ with same largest modulus, corresponding to the different subsets of type (ii) roots. Clearly, $\Delta=0$, so the circuit does not purify. 

\textbf{\emph{Interacting circuit: weakly purifying phase and purification transition}}---For $0<\gamma<\frac{\pi}{2}$, the model is interacting but integrable. We will show that in this case, the $\mathcal{A}$ symmetry breaking critical lines correspond to transitions between mixed and weakly purifying phases. In the weakly purifying phases, $\Delta\sim\frac{1}{L}$. We first study perturbatively away from the Gaussian limit, by setting $\gamma=\frac{\pi}{2}-\epsilon$, and then we study more general values of $\gamma$ by computing the Bethe roots numerically.

\onecolumngrid
\onecolumngrid
\vspace{\columnsep}
\begin{figure}[tb]
   \centering
   \includegraphics[width=0.98\columnwidth]{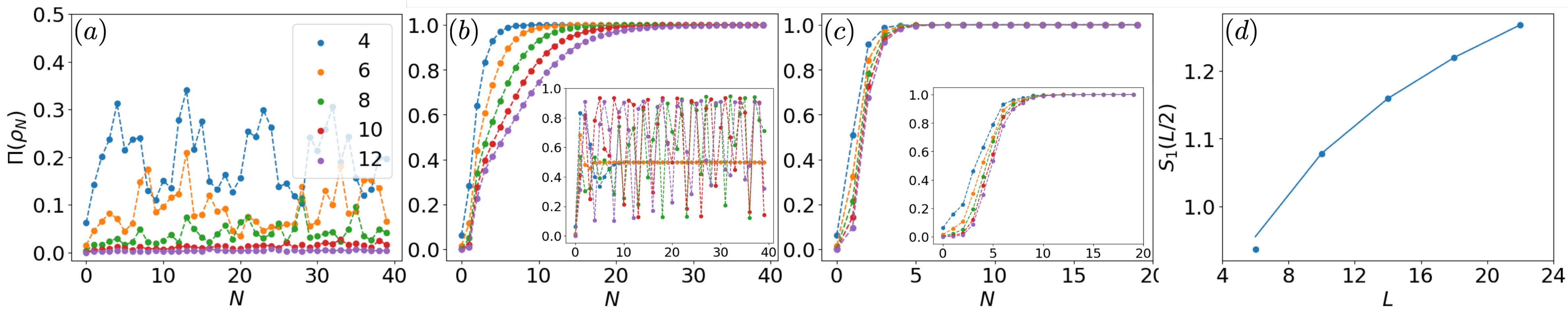} 
   \caption{$(a)$ The purity of $\rho_N$ in the symmetric (under $\mathcal{A}$) phase never converges to 1, meaning the circuit is mixed. For this plot, we used $\gamma=0.5$ and $T=1.0$. $(b)$ In the symmetry broken phase, the circuit is weakly purifying away from $\gamma=0,\frac{\pi}{2}$. Here, we use $\gamma=0.5$ and $T=\frac{\pi}{2\cos(0.5)}$. For $\gamma=0,\frac{\pi}{2}$, the circuit is mixed, and the latter is shown in the inset (with $T=1.5$). $(c)$ Upon adding integrability breaking perturbations in the symmetry breaking phase $(\gamma=0.5,\frac{\pi}{2\cos(0.5)},\delta=0.2$) or symmetry breaking perturbations in the symmetric phase $(\gamma=0.5,T=1.0,\delta'=0.2$), the circuit becomes stronlgy purifying. The latter is shown in the inset. $(d)$ The entanglement entropy in the weakly purifying phase demonstrates a log law. Here we use $\gamma=1.0, T=1.5$, and an initial state in the $+2$ magnetization sector. The line is a fit to $0.24\log L+0.52$.}
   \label{numerics}
\end{figure}
\vspace{\columnsep}

\twocolumngrid
First, let $\gamma=\frac{\pi}{2}-\epsilon$ with $\epsilon\ll 1$. For a given eigenvalue $\Lambda$ with a corresponding subset of roots $\mathcal{M}$, at first order in $\epsilon$ we find that the modification of $\log|\Lambda|$ is given by\cite{supp}
\begin{align}
\begin{split}
     &\partial_\epsilon \log|\Lambda|=-\frac{2}{L\tanh\beta}\\
     &\times\Im \sum_{m,n}(\tanh(2\lambda_m)-\tanh(2\lambda_n))\tanh(\lambda_m-\lambda_n)\,.
\end{split}
\end{align}

Pairs of roots of the same type do not modify $\log|\Lambda|$ at leading order. However, pairs $(\lambda_m,\lambda_n)$ of different types do modify $|\Lambda|$ at order $\epsilon$. Recall that $\mathcal{M}$ for the eigenvalues of $U_F$ with largest modulus at $\epsilon=0$ includes all the type (i) roots $\{\lambda_k^+\}$ with modulus smaller than $\frac{\pi}{4}$ and a subset of type (ii) roots. The exponentially large number of subsets of type (ii) roots for $\epsilon=0$ results in an exponential degeneracy in the eigenvalues with largest modulus. At first order in $\epsilon$, this degeneracy splits because each type (ii) root included in $\mathcal{M}$ modifies $|\Lambda|$. Specifically, a type (ii) root $\lambda_k^s=\mu\pm \frac{i\pi}{4}$, contributes to $\partial_{\epsilon}\log|\Lambda|$ as%Hence, since for $\epsilon=0$ all the type (i) roots smaller than $\pi/4$ in modulus have to be taken, at order $\epsilon$ each type (ii) root $\lambda_q=\nu\pm i\frac{\pi}{4}$ will contribute to $\partial_\epsilon \log|\Lambda|$ as
\begin{align}
\begin{split}
    &f_\pm(\mu)=\frac{4}{L\tanh\beta}\\
    &\times\sum_{\substack{k\\ \tan^2(\tfrac{2\pi k}{L})\sinh^2\beta<1}}\frac{-i\sinh(2\mu)\tanh(2\lambda_k^+)\mp \frac{\cosh(2\lambda_k^+)}{\tanh(2\mu)}}{\cosh(2\mu)\mp i\sin(2\lambda_k^+)}\,.
\end{split}
\end{align}
As a result, for finite $L$, there is a unique $\Lambda$ of largest modulus. This $\Lambda$ is given by $\mathcal{M}$ that includes all type (i) roots $\{\lambda_k^+\}$ with modulus smaller than $\frac{\pi}{4}$ and all type (ii) roots with real and imaginary part either both positive or both negative. Considering the smallest perturbation away from the largest $|\Lambda|$, we find that $\Delta\sim\frac{\epsilon}{L}$ \cite{supp}, and the circuit purifies in a time proportional to $\frac{L}{\epsilon}$. Due to the linear dependence of the purification time on $L$, the circuit is weakly purifying. 
% In the thermodynamic limit, this quantity becomes 
% \begin{equation}
%      f_\pm(\nu)=\frac{2}{\pi\tanh\beta\sinh\beta}\int_{-1}^1 dx \frac{\sinh(2\nu) \frac{x}{\sqrt{1-x^2}}\pm \frac{\sqrt{1-x^2}}{\tanh(2\nu)}}{(\cosh(2\nu)\mp x)(1+\tfrac{x^2}{\sinh^2\beta})}\,.
% \end{equation}

% In the thermodynamic limit, this gives
%\begin{equation}
 %    f_\pm(\nu)=\frac{2}{\pi\tanh\beta\sinh\beta}\int_{-1}^1 \D{x} \frac{\sinh(2\nu) \frac{x}{\sqrt{1-x^2}}\pm \frac{\sqrt{1-x^2}}{\tanh(2\nu)}}{(\cosh(2\nu)\mp x)(1+\tfrac{x^2}{\sinh^2\beta})}\,.
%\end{equation}
%This function vanishes as $\frac{k-L/4}{L}$ when $k\to \frac{L}{4}$. 
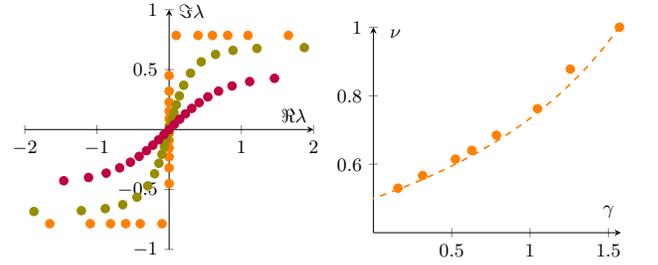
\begin{figure}
    \centering
   
\begin{tikzpicture}[scale=0.8]
\begin{axis}[
   scale=0.7, axis lines=middle,
    xmin=-2, xmax=2,
    ymin=-1, ymax=1.,
    xlabel={$\Re \lambda$},
    x label style={at={(axis description cs:1.,0.5)}},
    ylabel={$\Im \lambda$},
    y label style={at={(axis description cs:0.51,1.05)}},
    ]
]
\addplot [only marks,orange] table {
-1.6523097813335741727	-0.78539622374233509507
-1.0915965873398621157	-0.78539609253725601986
-0.81157642181629792018	-0.78539577846951247749
-0.60061691139930346277	-0.78539511227300453422
-0.40015841305172475674	-0.78539341079423144054
-0.097870190057259284017	-0.78537501304154441664
0	-0.45043839193241705397
0	-0.32002867610077081645
0	-0.22823974127215553905
0	-0.15413794815085739889
0	-0.089372623279546433794
0	-0.029309323320344627498
0	0.029309323320344627496
0	0.089372623279546433794
0	0.15413794815085739889
0	0.22823974127215553905
0	0.32002867610077081645
0	0.45043839193241705397
0.097870190057259284017	0.78537501304154441664
0.40015841305172475674	0.78539341079423144054
0.60061691139930346222	0.78539511227300453371
0.81157642181629792037	0.78539577846951247743
1.0915965873398621156	0.78539609253725601999
1.6523097813335741727	0.78539622374233509507
};

\addplot [only marks,olive] table {
-0.20164721422281899455	-0.36910695400670669261
-0.14200493780086565964	-0.28144949837679378428
-0.099816197036878114164	-0.20709826745957779527
-0.066684217822108413862	-0.14213698173986890294
-0.038404665142108599525	-0.083118921703717959046
-0.012554212511558322662	-0.027360499632950278862
0.012554212511558322663	0.027360499632950278863
0.038404665142108599526	0.083118921703717959048
0.066684217822108413868	0.14213698173986890294
0.099816197036878114189	0.20709826745957779527
0.14200493780086565968	0.28144949837679378427
0.20164721422281899457	0.3691069540067066926
-0.29575766205146742752	-0.46997958667375181783
-0.4446285569041113292	-0.56492689204595210835
-0.64246848906489856879	-0.6269002280665292134
-0.88417729800535072605	-0.65986819435463025638
-1.2156426481357901971	-0.67656753717993759011
-1.8729340714007627003	-0.68366433607376465042
0.29575766205146742753	0.46997958667375181783
0.4446285569041113292	0.56492689204595210835
0.64246848906489856879	0.6269002280665292134
0.88417729800535072605	0.65986819435463025638
1.2156426481357901971	0.67656753717993759011
1.8729340714007627003	0.68366433607376465042
};

\addplot [only marks,purple] table {
-0.53640251369739791916	-0.27250927662859114116
-0.41539577746436089758	-0.22405181239504098833
-0.31366898483342300656	-0.1764442467614802775
-0.22527748817809611119	-0.13040755864640924304
-0.14574216399916592099	-0.0859406440338447151
-0.0715970413934942385	-0.042658180795805653339
0	0
0.071597041393494238453	0.042658180795805653315
0.14574216399916592094	0.085940644033844715075
0.22527748817809611114	0.13040755864640924301
0.3136689848334230065	0.17644424676148027746
0.4153957774643608975	0.22405181239504098829
0.53640251369739791907	0.27250927662859114112
-0.68467356608545014447	-0.32001012569485271198
-0.87148630110986643712	-0.36365162715232205999
-1.1157083104592748406	-0.40018440132901511627
-1.4593087502990550807	-0.42713487105611450184
-2.0357127383034702051	-0.44338260756725616985
0.68467356608545014435	0.32001012569485271191
0.87148630110986643693	0.36365162715232205989
1.1157083104592748403	0.40018440132901511611
1.4593087502990550799	0.42713487105611450147
2.0357127383034702001	0.44338260756725616735
};
\end{axis}
\end{tikzpicture}
\begin{tikzpicture}[scale=0.8]
\begin{axis}[
    scale=0.6,axis lines=middle,
    xmin=0, xmax=1.58,
    ymin=0.4, ymax=1.,
    xlabel={$\gamma$},
    x label style={at={(axis description cs:1.,0.04)}},
    ylabel={$\nu$},
     y label style={at={(axis description cs:0.04,1.02)}}
    ]
]

\addplot [only marks,orange] table {
1.0472 0.762019
0.785398 0.684681
1.25664 0.8776
0.628319 0.640344
0.523599 0.614916 
0.314159 0.567027
0.15708 0.52996
1.5708 1
};

 \addplot[orange, dashed,thick,samples=100] {0.5/(1-x/3.14)};

\end{axis}
\end{tikzpicture}
    \caption{Left: Bethe roots in the complex plane corresponding to the largest modulus eigenvalue in size $L=48$, at $T=1.5$ and $\gamma=\frac{\pi}{2}$ (orange), $\gamma=1.04$ (green), $\gamma=0.52$ (purple). Right: Numerical estimation of the exponent $\nu$ as a function of $\gamma$. The dashed line is $\nu(\gamma)=\frac{1}{2(1-\gamma/\pi)}$. See SM \cite{supp} for details of the numerical study.}
    \label{betheroot}
\end{figure}

Far away from $\gamma=\frac{\pi}{2}$, we can no longer rely on the above perturbative result to obtain a $\Delta\sim\frac{1}{L}$. The Bethe ansatz solution allows for a nonperturbative study of the spectrum of the model for $\gamma\neq\frac{\pi}{2}$. From exact diagonalization in small systems, we find that the ground state of the model is unique and given by either $\frac{L}{2}$ or $\frac{L}{2}-1$ roots with a structure depicted in the left panel of Fig \ref{betheroot}. At large $L$, solving numerically the Bethe equations, we find that the solution with $\frac{L}{2}$ roots has always larger $|\Lambda|$, and that the state with $\frac{L}{2}-1$ roots has the second largest $|\Lambda|$. This structure is compatible with the ground state structure obtained at first order around $\gamma=\frac{\pi}{2}$. By looking at the difference between the two states with largest $|\Lambda|$, we find that the system of size $L$ purifies in a time $t_L=L\tau_L$ with some $\tau_L(\gamma,T)$ of order $1$ in the thermodynamic limit. Thanks to the Bethe ansatz, this purification time can be computed with excellent precision even in large system sizes. 
% This time can be computed exactly with the Bethe ansatz at the point $T=\frac{\pi}{2\cos\gamma}$
% \begin{equation}
%     \tau(\gamma,\tfrac{\pi }{2\cos\gamma})=\frac{\pi\tan(\frac{\pi^2}{4\gamma})}{\gamma(\pi-\gamma)}
% \end{equation}
% Away from this line, we have $\tau(\gamma,T)=f_{\gamma,\tau}(1/2)$ with $f_{\gamma,\tau}(m)$ a function whose expansion around $m=0$ can be computed explicitly. The first terms read
% \begin{equation}
%     f_{\gamma,\tau}(m)=
% \end{equation}

One interesting aspect of the mixed to weakly purifying transition is that it is well-defined even for finite system sizes. In particular, even for finite system sizes the purification time diverges as one approaches $T_c^-$ from above. %In finite size $L$, the purification time always diverges as $\tau_L(\gamma,T)\sim (T-T_c^-)^{-1}$ as $T\to T_c^+$, as shown in the SM. 
% In finite size $L$, the purification time diverges %as $\tau_L(\gamma,T)\sim (T-T_c^-)^{\nu_L}$ 
% as $T\to T_c^-$, with an exponent $\nu_L$ 
In the thermodynamic limit $L\to\infty$, we observe that the purification time diverges with an exponent $\nu$. Namely, by solving numerically the Bethe equations, we find that
\begin{equation}
    \tau_{L\to\infty}(\gamma,T)\sim (T-T_c^-)^{-\nu(\gamma)}\,,
\end{equation}
where $\nu(\gamma)$ has a dependence on $\gamma$ shown numerically in the right panel of Fig \ref{betheroot}. The numerical results fit well with the analytic function $\nu(\gamma)=\frac{1}{2(1-\gamma/\pi)}$. For generic $\gamma$, we show in the supplemental material\cite{supp} that when $T\to T_c^-$, the Bethe equations for the ground state split into two sets of decoupled XXZ Bethe equations, which suggests a 
relation between the observed critical behaviour and that of the XXZ spin chain. We note that $\frac{1}{1-\gamma/\pi}$ is indeed a critical exponent of the XXZ chain \cite{luther1975,lukyanov1999correlation}.

\textbf{\emph{Integrability breaking and symmetry breaking.}}---We now generalize the model (\ref{model1}) in two ways. First, we add a perturbation that respects $\mathcal{A}$ but breaks integrability. We define 
\begin{equation}
U_F'=U_3U_FU_3\qquad U_3=\prod_m\mathrm{exp}(\delta (-1)^m\sigma^z_m)\,.
\end{equation}
It is straightforward to check that $U_F'$ still respects $\mathcal{A}$, but is no longer integrable. Once $\delta\gg \frac{1}{L}$, the circuit becomes strongly purifying in the symmetry broken phase, because the integrability-breaking perturbation splits the eigenvalues of $U_F$ by $\mathcal{O}(\delta)$. Therefore, for large $L$, the purification rate should be independent of $L$, but for small $L$, the circuit may still appear weakly purifying. We compute $\Pi(\rho_N)$ numerically, as shown in Fig~\ref{numerics}.(c), to confirm this result.%and find that it turns regions of the phase diagram that were previously weakly purifying into strongly purifying. This is shown in Fig.~[...]. 

Next, we add a perturbation that preserves integrability, but explicitly breaks the antiunitary symmetry. We define
\begin{equation}
U_F'=U_3'U_F\qquad U_3'=\prod_m\mathrm{exp}(\delta'\sigma^z_m)\,.
\end{equation}
This perturbation removes the volume-law phase, because it splits the uni-modular eigenvalues of $U_F$ and projects onto the unique ferromagnetic eigenstate with largest eigenvalue under $U_3'$ (see the inset of Fig.~\ref{numerics}.(c)).

\textbf{\emph{Discussion.}}---We show through a family of simple models that there is a hierarchy of different purification transitions in non-unitary circuits, depending on whether the circuit is Gaussian, integrable, or non-integrable. One important next step would be to understand whether the weakly purifying phase is generic to interacting but integrable systems. In other words, do antiunitary symmetry breaking transitions in interacting, integrable circuits always correspond to mixed to weakly purifying transitions? It would also be interesting to determine the critical exponents for the divergence of the purification time analytically, and confirm explicitly the relation between these exponents and the magnetization exponent in the stationary Heisenberg XXZ model. Finally, the behavior of the entanglement entropy, and its dependence on the magnetization sectors of the initial state, remain to be better analytically understood.  

\textbf{\emph{Acknowledgements}}---C.Z. acknowledges support from the University of Chicago Bloomenthal Fellowship and the National
Science Foundation Graduate Research Fellowship under Grant No. 1746045, and the Harvard Society of Fellows. E.G. acknowledges support from the Kadanoff Center for Theoretical Physics at University of Chicago, and from the Simons Collaboration on Ultra-Quantum Matter.
%%%%%%%%%%%%%%%%%%%%%%%%%%%
%%%%%%%%%%%%%%%%%%%%%%%%%%%%%%%%%%%%%%%%%%%%%%%%%%%%%%%%%
\bibliography{bibliography}
%\bibliographystyle{plain}
%%%%%%%%%%%%%%%%%%%%%%%%%%%%%%%%%%%%%%%%%%%%%%%%%%%%%%%%%%%%%%%%%%%%%%%%%%%%%%%%%%%%

\end{document}

\title{Supplemental Material}
\author{Carolyn Zhang}
\affiliation{Department of Physics, Kadanoff Center for Theoretical Physics, University of Chicago, Chicago, Illinois 60637,  USA}
\affiliation{Department of Physics, Harvard University, Cambridge, MA02138, USA}
\author{Etienne Granet}
\affiliation{Quantinuum, Leopoldstrasse 180, 80804 Munich, Germany}
\date{\today}

\maketitle
\section{Gaussian circuit}
Here we present a more detailed analysis of the integrable circuit at $\gamma=\frac{\pi}{2}$, where the circuit is Gaussian at all values of $T$. For this particular value of $\gamma$, we have
\begin{equation}
    \alpha=-\frac{1}{2}\log \frac{1+T}{1-T}.
\end{equation}
Plugging this into the Bethe equations (Eq. 14 of the main text), we get
\begin{equation}
    \left( \frac{\sinh(\lambda_m+\alpha/2+i\pi/4)\sinh(\lambda_m-\alpha/2+i\pi/4)}{\sinh(\lambda_m+\alpha/2-i\pi/4)\sinh(\lambda_m-\alpha/2-i\pi/4)}\right)^{L/2}=(-1)^{M-1}\,.
\end{equation}
Therefore,
\begin{equation}
     \frac{\sinh(\lambda_m+\alpha/2+i\pi/4)\sinh(\lambda_m-\alpha/2+i\pi/4)}{\sinh(\lambda_m+\alpha/2-i\pi/4)\sinh(\lambda_m-\alpha/2-i\pi/4)}=e^{\frac{4i\pi k}{L}}\,,
\end{equation}
with $-\frac{L}{4}<k<\frac{L}{4}$ integer/half-integer if $M$ is odd/even. The solutions are then given by
\begin{equation}
    e^{2\lambda_k^\pm}=-\tan(\tfrac{2\pi k}{L})\cosh\alpha\pm\sqrt{\tan(\tfrac{2\pi k}{L})^2\cosh\alpha^2+1}\,.
\end{equation}
The eigenvalues are thus parameterized by a subset $\mathcal{M}=\{\lambda_k^s\}$ of these solutions, where $s\in\{+,-\}$. They take the explicit form
\begin{equation}\label{lambdaeq}
    \Lambda=\prod_{\lambda_k^s\in\mathcal{M}} \frac{\cosh(2\lambda_k^s)-i\sinh\alpha}{\cosh(2\lambda_k^s)+i\sinh\alpha}\,.
\end{equation}
When $0<T\leq1$, $\alpha$ is real and all the eigenvalues have modulus $1$, so the antiunitary symmetry $\mathcal{A}$ is not broken. When $T>1$, $\mathcal{A}$ is broken and the eigenvalues $\Lambda$ do not all have modulus $1$. Let us consider the case $T>1$ and look for eigenvalues with largest modulus. Denoting
\begin{equation}
    \beta=-\frac{1}{2}\log \frac{T+1}{T-1}=\alpha-i\frac{\pi}{2}\,,
\end{equation}
we have
\begin{equation}
    e^{2\lambda_k^\pm}=-i\tan(\tfrac{2\pi k}{L})\sinh\beta\pm\sqrt{1-\tan^2(\tfrac{2\pi k}{L})\sinh^2\beta}\,,
\end{equation}
so
\begin{equation}
    \cosh(2\lambda_k^\pm)=\pm\sqrt{1-\tan^2(\tfrac{2\pi k}{L})\sinh^2\beta}\,.
\end{equation}

In terms of $\beta$, the eigenvalues $\Lambda$ take the form
\begin{equation}
    \Lambda=\prod_{\lambda_k^s\in\mathcal{M}} \frac{\cosh(2\lambda_k^s)+\cosh\beta}{\cosh(2\lambda_k^s)-\cosh\beta}\,.
\end{equation}
When $\tan(\tfrac{2\pi k}{L})^2\sinh^2\beta>1$, $\cosh(2\lambda^\pm_k)$ is purely imaginary and $\lambda^\pm_k$ does not contribute to $|\Lambda|$. On the other hand, when $\tan(\tfrac{2\pi k}{L})^2\sinh^2\beta<1$ the $\lambda_k^+$ increases $|\Lambda|$. It follows that the largest $|\Lambda|$'s are obtained by taking all roots $\lambda_k^+$ such that $\tan(\tfrac{2\pi k}{L})^2\sinh^2\beta<1$, and any subset of $\{\lambda_k^s\}$ such that $\tan(\tfrac{2\pi k}{L})^2\sinh^2\beta>1$ (where $s$ can be either $+$ or $-$). The existence of the roots with $\tan(\tfrac{2\pi k}{L})^2\sinh^2\beta>1$ means that for any $\beta>0$, i.e. any $T>1$, there are an exponentially large number of eigenvalues with same largest modulus. Hence we have a mixed phase for any $T>0$.\\

Let us briefly comment on the structure of the Bethe roots in this free fermion limit. When $\tan(\tfrac{2\pi k}{L})^2\sinh^2\beta<1$, $e^{2\lambda_k^s}$ is of modulus $1$ and so $\lambda_k^s$ is purely imaginary. Its imaginary part is between $\pm \frac{\pi}{4}$ when $s=+$, and is larger when $s=-$. When  $\tan(\tfrac{2\pi k}{L})^2\sinh^2\beta>1$, $e^{2\lambda_k^s}$ is purely imaginary and so $\lambda_k^s$ has imaginary part exactly $\pm \frac{\pi}{4}$. So the largest modulus eigenvalues are given by Bethe roots filling the imaginary axis between $\pm \frac{i\pi}{4}$, and arbitrary sets of roots with imaginary part $\pm\frac{\pi}{4}$ with non-zero real part.

\section{First order in $\epsilon$ at fixed $T$}
We now perturb away from the $\gamma=\frac{\pi}{2}$ line at fixed $T$. We set
\begin{equation}
    \gamma=\frac{\pi}{2}-\epsilon.
\end{equation}

Let us introduce
\begin{equation}
    s(\lambda_m)=-i\log \frac{\sinh(\lambda_m+\alpha/2+i\pi/4)\sinh(\lambda_m-\alpha/2+i\pi/4)}{\sinh(\lambda_m+\alpha/2-i\pi/4)\sinh(\lambda_m-\alpha/2-i\pi/4)}\,.
\end{equation}
At order $\epsilon$, the Bethe equations in logarithmic form are
\begin{equation}
    s(\lambda_m)-\epsilon(\tanh(2\lambda_m+\alpha)+\tanh(2\lambda_m-\alpha))=\frac{4\pi I_m}{L}-\frac{4\epsilon}{L}\sum_{n\neq m}\tanh(\lambda_m-\lambda_n)\,.
\end{equation}

The numbers $I_m$ are either all integers when the number of roots $M$ is odd, and all half-integers when $M$ is even.  We have
\begin{equation}
    s'(\lambda_m)=-\frac{2}{\cosh(2\lambda_m+\alpha)}-\frac{2}{\cosh(2\lambda_m-\alpha)}\,.
\end{equation}
Expanding the Bethe roots about the $\epsilon=0$ point, we get $\lambda_m=\lambda_m(\epsilon=0)+\epsilon\mu_m$ with 
\begin{equation}
    \mu_m=\frac{-\tanh(2\lambda_m+\alpha)-\tanh(2\lambda_m-\alpha)+\frac{4}{L}\sum_{n\neq m}\tanh(\lambda_m-\lambda_n)}{\frac{2}{\cosh(2\lambda_m+\alpha)}+\frac{2}{\cosh(2\lambda_m-\alpha)}}\,.
\end{equation}
It follows that
\begin{equation}
    \partial_\epsilon \log\Lambda=\sum_{m} 2i\left[\frac{1}{\cosh(2\lambda_m-\alpha)}-\frac{1}{\cosh(2\lambda_m+\alpha)} \right]\mu_m+i(\tanh(2\lambda_m+\alpha)-\tanh(2\lambda_m-\alpha))\,,
\end{equation}
and
\begin{equation}
     \partial_\epsilon \log|\Lambda|=-\frac{2}{L\tanh\beta}\Im \sum_{m,n}(\tanh(2\lambda_m)-\tanh(2\lambda_n))\tanh(\lambda_m-\lambda_n)\,.
\end{equation}
Pairs of  $\lambda_m,\lambda_n$ that are either both purely imaginary or both complex with imaginary part $\pm \frac{i\pi}{4}$ do not modify $|\Lambda|$ at leading order. However, pairs $(\lambda_m,\lambda_n)$ where $\lambda_m$ is purely imaginary and $\lambda_n$ is complex with imaginary part $\pm \frac{i\pi}{4}$ will modify $|\Lambda|$. Since for $\epsilon=0$ the largest $|\Lambda|$s are obtained by taking the roots $\{\lambda_k^+\}$ with $\tan(\tfrac{2\pi k}{L})^2\sinh^2\beta<1$, at order $\epsilon$ each $\lambda_q^s=\mu\pm i\frac{\pi}{4}$ with $\tan(\tfrac{2\pi q}{L})^2\sinh^2\beta>1$ will contribute to $\partial_\epsilon \log|\Lambda|$ as
\begin{equation}
    f_\pm(\mu)=\frac{4}{L\tanh\beta}\sum_{\substack{k\\ \tan^2(\tfrac{2\pi k}{L})\sinh^2\beta<1}}\frac{-i\sinh(2\mu)\tanh(2\lambda_k^+)\mp \frac{\cosh(2\lambda_k^+)}{\tanh(2\mu)}}{\cosh(2\mu)\mp i \sinh(2\lambda_k^+)}\,.
\end{equation}
In the thermodynamic limit this gives
\begin{equation}
     f_\pm(\mu)=\frac{2}{\pi\tanh\beta\sinh\beta}\int_{-1}^1 \D{x} \frac{\sinh(2\mu) \frac{x}{\sqrt{1-x^2}}\mp \frac{\sqrt{1-x^2}}{\tanh(2\mu)}}{(\cosh(2\mu)\pm x)(1+\tfrac{x^2}{\sinh^2\beta})}\,.
\end{equation}
This is non-zero for finite $\mu$, so the exponentially large degeneracy in finite size is lifted at first order in $\epsilon$. However, in terms of the quantized values of $\mu$ (through the quantized values of $\lambda_k^s$), this function vanishes as $(\frac{L}{4}-k)/L$ when $k\to \frac{L}{4}$. Therefore, $\Delta\sim\frac{1}{L}$. %Hence at first order in $\epsilon$ there is a degeneracy in the thermodynamic limit, roughly given by the partitions of integers $L^\nu$ for any $\nu<1$, so asymptotically $\sim e^{\sqrt{L}}$.

\section{Limit $T\to T_c^-$}
When $T\to T_c^-$ from above, we have $\beta\to-\infty$. %Let us consider a set of $\frac{L}{2}$ roots $\lambda_j$ that are symmetric around $0$. For the roots with positive real part, 
We introduce $\mu_j$ by
\begin{equation}
    \lambda_j= \frac{\alpha}{2}+\mu_j\,.
\end{equation}
From the numerics, we look for a solution such that the $\mu_j$'s have a finite limit when $\beta\to-\infty$. Then we have
\begin{equation}
    \left[ \frac{\sinh(\mu_i+i\gamma/2)}{\sinh(\mu_i-i\gamma/2)}\right]^{L/2}=e^{iL\gamma}\prod_{j\neq i}\frac{\sinh(\mu_i-\mu_j+i\gamma)}{\sinh(\mu_i-\mu_j-i\gamma)}\,,
\end{equation}
which are the Bethe ansatz equations of the XXZ chain on $\frac{L}{2}$ sites with a twist $e^{iL\gamma}$. In this limit, the solutions $\mu_j$ are either real or come in complex conjugate pairs, and the eigenvalue $\Lambda$ becomes
\begin{equation}
\Lambda=\left[\prod_{j=1}^{L/2}\frac{\sinh(\mu_j-i\gamma/2)}{\sinh(\mu_j+i\gamma/2)}\right]^2\,,
\end{equation}
which are of modulus $1$ when the $\mu_j$'s are real or come in complex conjugate pairs.

\section{Numerical study of the critical behaviour near $T\to T_c^+$}
We now detail the numerical study using the Bethe ansatz of the critical behaviour of the purification time near $T_c^+$, given by
\begin{equation}\label{scaling}
    \tau_\infty(\gamma,T)\sim (T-T_c)^{-\nu}\,.
\end{equation}
We start by converting the Bethe equations into their logarithmic form by applying $\log$ on both sides. The logarithmic form reads
\begin{equation}\label{logbethe}
    s(\lambda_i)=\frac{I_i}{L}+\frac{1}{L}\sum_{j=1}^M r(\lambda_i-\lambda_j)\,,
\end{equation}
with
\begin{equation}
    r(\lambda)=\frac{1}{\pi}\arctan \frac{\tanh \lambda}{\tan\gamma}\,,
\end{equation}
and
\begin{equation}
    s(\lambda)=-\frac{1}{2\pi}\arctan \frac{\tan (\gamma/2)}{\tanh(\lambda+\alpha/2)}-\frac{1}{2\pi}\arctan \frac{\tan (\gamma/2)}{\tanh(\lambda-\alpha/2)}\,.
\end{equation}
Under this form, we find that the state with largest $|\Lambda|$ for $L$ sufficiently large is given by $M=\frac{L}{2}$ numbers $I_i=-\frac{M-1}{2},...,\frac{M-1}{2}$. We also find that the state with second largest $|\Lambda|$ is given by $M=\frac{L}{2}-1$ numbers $I_i=-\frac{M-1}{2},...,\frac{M-1}{2}$, i.e. one root less than for the largest, but still fully packed around the origin. To find this structure, we used exact diagonalisation in small sizes up to $L\sim 20$, for which the largest modulus eigenvalue and second largest modulus eigenvalue are always of the previous form, except sometimes swapped, namely the state with largest $|\Lambda|$ is obtained with $\frac{L}{2}-1$ roots. However, using the Bethe equations to compute the largest $|\Lambda|$ for larger system sizes $L$, we find that the largest $|\Lambda|$ always ends up being given by $\frac{L}{2}$ roots for $L$ sufficiently large (i.e. greater than $\sim 50$).

To determine the scaling \eqref{scaling} from these root structures, we proceed as follows. The purification time in finite size is $L \tau_L(\gamma,T)=1/\log \frac{|\Lambda_0|}{|\Lambda_1|}$ with $\Lambda_{0,1}$ corresponding to the largest and second largest $|\Lambda|$. At fixed $\gamma$ and $T$, we solve numerically \eqref{logbethe} by means of a Newton method for increasingly large $L$, by using the solution in size $L$ as a seed for the initial guess of size $L+2$. Specifically, we slightly shift all the roots of the solution at $L$ with positive real part and add one additional root at zero. In this way we proceed up to size $288$. This kind of technique to solve Bethe equations for large system sizes is usual, see e.g. \cite{vernier2014non}. The value $\log |\Lambda_{0}|/|\Lambda_{1}|$ behaves as $1/L$ with corrections $1/L^2$. We use thus a linear fit in $1/L$ with the sizes studied to extrapolate the value of $\tau_\infty(\gamma,T)$ with excellent precision. Then, at fixed $\gamma$, we progressively decrease $T$ towards $T_c^+$. Using values of $T-T_c^+=0.0125,0.00625,0.003125,0.0015625$ we deduce the value of $\nu(\gamma)$ by doing a linear fit in logarithmic plot.

%%%%%%%%%%%%%%%%%%%%%%%%%%%%%%%%%%%%%%%%%%%%%%%%%%%%%%%%%%%%%%%%%%%%%%%%%%%%%%%%%%%%
\bibliography{bibliography}
%%%%%%%%%%%%%%%%%%%%%%%%%%%%%%%%%%%%%%%%%%%%%%%%%%%%%%%%%%%%%%%%%%%%%%%%%%%%%%%%%%%%